\documentclass[12pt,reqno]{amsart}
\usepackage{amsmath}
\usepackage{amssymb}
\usepackage{amscd}
\usepackage{mathrsfs}
\usepackage{amsthm}

\theoremstyle{definition}
\newtheorem{definition}{Definition}
\theoremstyle{definition}
\newtheorem{example}{Example}
\theoremstyle{definition}
\newtheorem{remark}{Remark}
\theoremstyle{plain}
\newtheorem{theorem}{Theorem}

\theoremstyle{plain}

\begin{document}

 \title[Differential invariants of measurements]{Differential invariants of measurements, and their connection to central moments}
 \author[E. Schneider]{Eivind Schneider}
 \date{}
\address{
Faculty of Science, University of Hradec Králové, Rokitanskeho 62, Hradec Králové 50003, Czech Republic. \newline
E-mail address: {\tt eivind.schneider@uhk.cz}. }
 \keywords{Statistical mechanics, information theory, measurement, thermodynamics, symplectic geometry, contact geometry, central moments, differential invariants, heat capacity.}

 \begin{abstract}
Due to the principle of minimal information gain, the measurement of points in an affine space $V$ determines a Legendrian submanifold of $V \times V^* \times \mathbb R$. Such Legendrian submanifolds are equipped with additional geometric structures that come from the central moments of the underlying probability distributions and are invariant under the action of the group of affine transformations on $V$.

We investigate the action of this group of affine transformations on Legendrian submanifolds of $V \times V^* \times \mathbb R$ by giving a detailed overview of the structure of the algebra of scalar differential invariants. We show how the central moments can be used to construct the scalar differential invariants. In the end, we view the results in the context of equilibrium thermodynamics of gases, where we notice that the heat capacity is one of the differential invariants. 
 \end{abstract}

 \maketitle


\section{Introduction}
Already in \cite{Gibbs1,Gibbs2} Gibbs knew that thermodynamics can be formulated in the language of contact geometry.  Since the first law of thermodynamics takes the form of a contact structure on an odd-dimensional manifold, thermodynamic states correspond to Legendrian submanifolds with respect to this contact structure. 
More recently, some effort has been devoted towards studying an additional metric structure appearing on these Legendrian submanifolds (see for example \cite{RuppeinerMetric,RuppeinerFluctuation} and references therein). 

Both the contact structure and the metric can be interpreted as coming from information theory and statistical mechanics. If we model the process of measuring a physical quantity  by a random vector in an affine space $V$, the principle of minimal information gain (analogue to Jaynes' principle of maximal entropy, \cite{Jaynes1}) gives a contact structure on $V \times V^* \times \mathbb R$. Thus, a measurement process is in this sense always accompanied by a theory of ``thermodynamics'' (\cite{Jaynes1,MNSS,Measurement}). The statistical viewpoint gives additional geometric structures on Legendrian submanifolds. The variance takes the form of a metric on Legendrian submanifolds (\cite{MNSS}). In fact, for each integer $k \geq 2$, the $k$th central moment of gives a symmetric $k$-form on Legendrian submanifolds (\cite{Measurement}). 

In this framework the group of affine transformations of $V$ appears naturally as symmetries of the structures appearing, and defines an equivalence relation on the space of Legendrian submanifolds. Taking this Lie group action as a fundamental part of the theory, it becomes clear that the important quantities in  theory are those that are invariant under this group action. These quantities will be the main focus of this paper. In particular, we will give a detailed description of the algebra of scalar differential invariants, building upon the work done in \cite{Schneider}.

We will apply our results to the thermodynamics of gases. In particular we notice that the simplest scalar differential invariant corresponds to the heat capacity in thermodynamics. 


We start in Section \ref{FromStatisticsToGeometry} by recalling how the process of measuring vectors leads to symplectic and contact geometry, following \cite{Measurement}. We explain how Legendrian submanifolds come equipped with symmetric $k$-forms corresponding to $k$th central moments. We end the section by discussing the action of the Lie group  $\mathrm{Aff}(V)$ of affine transformations on $V$. 

In Section \ref{Jets} we recall notions from the geometric theory of PDEs and explain the concept of scalar differential invariants. We show how $\mathrm{Aff}(V)$ acts on the jet spaces, and that it is the largest group acting on $V$ that preserves the central moments. 

In Section \ref{Invariants} we use the central moments to find generators for the field of scalar differential invariants. We compute the Hilbert and Poincaré functions for the field.  We also find invariant derivations, and give a finite set of generators for the set of differential invariants, as a differential algebra. We compute differential syzygies for the case $\dim V=2$. 

In Section \ref{Gases} we discuss the results from the previous section in the context of gases. We find a generating set of differential invariants and invariant derivations with respect to the group of affine transformations, and with respect to a two-dimensional subgroup.

\section{From random vectors to differential geometry}\label{FromStatisticsToGeometry}
We start by describing how the principle of minimal information gain applied to the measurement of points in an affine space leads to contact and symplectic geometry. We follow closely the approach of \cite{Measurement}. 

\subsection{Measuring vectors}
A random vector is a map from a probability space to the space $V= \mathbb R^n$: 
\[X \colon (\Omega,\mathcal A,\mu_0) \to V\] 
Here $\Omega$ is the sample space, $\mathcal A$ the $\sigma$-algebra of ``events'', and $\mu_0$ a probability measure. We interpret $X$ as a measurement of $x_0 \in V$ if \[E_{\mu_0}(X)=\int_\Omega X d\mu_0=x_0.\] 
By choosing a basis in $V$, the integral above can be defined coordinate-wise. 

A different vector $x \in V$ can be measured by changing the probability measure from $\mu_0$ to $\mu$, using the probability measure as a control parameter, in a way so that $E_\mu(X)=x$. We find conditions that determine $\mu$. 

Assume first that $\mu_0$ is absolutely continuous with respect to $\mu$. Then the Radon-Nikodym theorem implies that $d\mu=\rho d\mu_0$. We require $\rho$ to satisfy the conditions
\begin{equation}
E(X_{\mu})=\int_\Omega X \rho d\mu_0=x, \qquad \int_\Omega \rho d\mu_0 = 1. \label{rho1}
\end{equation}
In addition we require $\mu$ to be the probability distribution closest to $\mu_0$ in a certain sense: We require it to minimize the information gain, or Kullback-Leibler divergence:
\[I(\mu,\mu_0)= \int_\Omega \rho \ln \rho \, d\mu_0 \]
This is the principle of minimal information gain. It is analogous to Jaynes' maximum-entropy principle (\cite{Jaynes1}). As Jaynes put it, it is the only unbiased assignment we can make. ``To use any other would amount to arbitrary assumption of information which by hypothesis we do not have.''  

Note that Jensen's inequality gives $I(\mu,\mu_0)\geq 0$. Moreover, we have $I(\mu,\mu_0)=0$ if $\rho=1$. 

Minimizing $I(\mu,\mu_0)$ gives  
\begin{equation}
 \rho=\frac{1}{Z(\lambda)} e^{\langle \lambda, X\rangle} \label{rho2}
 \end{equation}
where $\lambda \in V^*$ and $Z(\lambda)=\int_\Omega e^{\langle \lambda, X \rangle} d\mu_0$ is the {partition function} (see \cite{Measurement} for details). Thus $\rho$ is determined by $\lambda$, and $\lambda=0$ gives $\rho=1$. 

Let $D \subset V^*$ be a {simply connected} 
 domain, containing $0$, on which $Z(\lambda)$ is defined and smooth.  Due to the first condition of (\ref{rho1}), we have $d_\lambda Z=Z(\lambda)x$. (The differential $d_\lambda Z$ is an element in $T^*_\lambda D$ which can be identified with $V=(V^*)^*$ by using the affine structure on $V^*$.) By defining $H(\lambda)=-\ln Z(\lambda)$, we get
\begin{equation}
x=-d_\lambda H. \label{x}
\end{equation} 
This defines an $n$-dimensional manifold \[L=\{x=-d_\lambda H \mid \lambda \in D\} \subset V \times V^*. \] 
By choosing a basis on $V$ (and dual basis on $V^*$), we get coordinates $x^i$ on $V$ and $\lambda_i$ on $V^*$. In these coordinates $L$ is given by $n$ equations $x^i=-H_{\lambda_i}$. 

The space $V \times V^*$ is a symplectic space equipped with the symplectic form 
\begin{equation}
\omega = d \lambda_i \wedge d x^i, \label{eq:omega}
\end{equation} 
and $L$ is a Lagrangian submanifold, i.e. $\omega|_L=0$. We use the Einstein summation convention, and sum over repeated indices.


Since the information gain $I(\mu,\mu_0)$ depends on $\rho$, and therefore on $\lambda$, it can be considered as a function on $L$. By using (\ref{rho2}) we get \[I(\mu,\mu_0)=H(\lambda)+\langle \lambda, x \rangle =H(\lambda)-\langle \lambda , d_\lambda H\rangle\] on $L$.  
If $x\in V$ can be measured, then equation (\ref{x}) can be solved for $\lambda$. If $\lambda(x)$ is a (local) solution, we may write
\[I(x)=H(\lambda(x))+\langle \lambda(x),x)\rangle.\]  
Let $u$ be a coordinate on $\mathbb R$. We have the relation $I_{x^i} = \lambda_i$, so the submanifold
\[\tilde L = \{u=I(x), \lambda_i=I_{x^i}(x)\} \subset V \times V^* \times \mathbb R\] 
is Legendrian with respect to the contact form $\theta = du-\lambda_i dx^i$ on $V \times V^* \times \mathbb R$. 

\begin{remark}
The solution $\lambda(x)$ will in general be multivalued, where each value correspond to different phases of the system. In this sense, the description of $\tilde L$ above is just a local one, and the local parts should be considered together as one Legendrian submanifold.
\end{remark}

The Legendrian manifold $\tilde L$ depends on the initial distribution $\mu_0$. In this way, the statistical object we started with is translated into a geometric object $\tilde L$. This shows that our model of physical measurements, based on random vectors and the principle of minimal information gain, is always accompanied by a form of thermodynamics, where the contact form $du-\lambda_i dx^i$ plays the role of the first law.  In this interpretation, the Legendrian submanifolds of $V \times V^* \times \mathbb R$ obtained in this way correspond to thermodynamic states.

\subsection{Central moments}

Let $\tilde L$ continue to denote the Legendrian manifold corresponding to a measurement modelled on the random vector $X$ considered above. It can be parametrized either by the measured quantity $x \in V$ or by the parameter $\lambda \in V^*$. 

The notion of a metric on  $\tilde L$ has received a great deal of attention in the context of thermodynamics. In the framework presented above, one metric naturally appears, namely the one which is defined as the variance of the random vector $X$ (with respect to the extremal measure $\mu$ given by $\rho$, or $\lambda$). This is only one of an infinite number of symmetric forms on $\tilde L$, called central moments.

The $k$th moment of $X$ (with respect to $\mu$) is defined by 
\[m_k= \int_\Omega X^{\otimes k} \rho d\mu_0 \in S^kV \subset V^{\otimes k}. \] 
It depends on $\lambda$ and defines a symmetric tensor on $\tilde L$. If we choose coordinates $\lambda_i$ on $\tilde L$, it can be written as 
\begin{align*}
 m_k &= \left(\int_\Omega X^{i_1}  \cdots  X^{i_k} \rho d \mu_0 \right) d \lambda_{i_1} \otimes \cdots \otimes d \lambda_{i_k} \\
 &=  \frac{Z_{\lambda_{i_1} \cdots \lambda_{i_k}}}{Z} d \lambda_{i_1} \otimes \cdots \otimes d \lambda_{i_k}. 
\end{align*}
Assuming that the function $Z$ is smooth, the differentiation is symmetric, and thus $m_k$ is a symmetric $k$-form.
The equality $Z \int_\Omega X^\tau \rho d \mu_0 =Z_{\lambda_\tau}$ can be shown by induction on $\tau$ using (\ref{rho2}). The symmetric $k$-form $m_k$ is $GL(V)$-invariant. 

We define the $k$th central moment $\sigma_k$ of $X$ as the $k$th moment of $X-m_1(X)$. The central moment $\sigma_k$ is related to $m_k$ through the formula
\[\sigma_k = \sum_{i=0}^k (-1)^{k-i}\binom{k}{i} m_i \odot m_1^{\otimes (k-i)}.\] 
The central moments are invariant under the action of the  group $\mathrm{Aff}(V)$ of affine transformations on $V$.

Note that each central moment $\sigma_k$ is completely determined by $\tilde L$, or even by the corresponding Lagrangian manifold $L \subset V \times V^*$.

As mentioned, the second central moment 
\[\sigma_2=-H_{\lambda_i \lambda_j} d\lambda_i \otimes d\lambda_j\] gives a positive definite metric on Legendrian submanifolds. It can also be given locally as $I_{x^i x^j} dx^i \otimes dx^j$. Or if we consider only the Lagrangian submanifold $L \subset V \times V^*$, given by $n$ functions $x^i(\lambda)$, it can be given by $x^i_{\lambda_j}d\lambda_i \otimes d\lambda_j$. Notice that the positive definiteness of $\sigma_2$ puts conditions on $\tilde L$. Thus the measurement does not lead to arbitrary Legendrian manifolds, but only to those where the symmetric $k$-forms can be properly interpreted as central moments. 

The metric $\sigma_2$ have been studied previously in different contexts. In information geometry $\sigma_2$ is treated as a metric on $V^*$ (or on the space of probability measures), where it is called the Fisher information metric. In geometric thermodynamics it was treated by Ruppeiner (\cite{RuppeinerMetric}), and the statistical interpretation of Ruppeiner's metric was made by \cite{MNSS}. Ruppeiner was also interested in computing and giving a physical interpretation to its Riemannian curvature. 
However, as was pointed out in \cite{Measurement}, and which we will discuss further in this paper, the group  $\mathrm{Aff}(V)$ which appears naturally in this context, and acts on Legendrian submanifolds, is smaller than the full Lie pseudogroup of diffeomorphisms. As a result, the Riemannian curvature is not the most fundamental invariant.

\begin{remark}[\cite{MNSS}]
The metric $\sigma_2$ can be extended to a metric on $V \times V^*$ or $V  \times  V^*\times \mathbb R$. 
On $V \times V^*$ we can define a quadratic form 
\[\kappa = \frac{1}{2} \sum_{i=1}^n (d\lambda_i \otimes dx^i+dx^i \otimes d\lambda_i)\]
which clearly restricts to $\sigma_2$ on a Lagrangian submanifold.  
To get a nondegenerate metric on $V \times  V^*\times \mathbb R$, we may add a multiple of $(du-\lambda_i dx^i)^2$.
\end{remark}

\subsection{Action of the group of affine transformations} \label{Group}
From the statistical approach outlined above, we get naturally an action of the affine group $\mathrm{Aff}(V)$. One can view it as coming from the arbitrary choice of origin and basis in $V$. The action extends to $V \times V^* \times \mathbb R$ in the obvious way: 
\[(x,u,\lambda) \mapsto (A x+B,u,(A^{-1})^T \lambda), \qquad A \in GL(V), B \in V \]
This Lie group action preserves both the contact structure on $V \times V^* \times \mathbb R$ (or symplectic structure on $V \times V^*$) and the central moments. 
We will show in Theorem \ref{unique} in Section \ref{Jets} that $\mathrm{Aff}(V)$ is uniquely determined as the largest Lie group acting on $V$ which preserves all of these structures.

This insight gives $\mathrm{Aff}(V)$ a special place in the theory of measurements, which was pointed out in \cite{Measurement}, and this Lie group will be the focus of our attention in most of this paper. 

The $\mathrm{Aff}(V)$-action on $V \times V^* \times \mathbb R$ induces an action on Legendrian submanifolds. If we treat Legendrian submanifolds that are related by such a transformation as equivalent, it becomes clear that the important geometric structures on Legendrian submanifolds are those that are invariant under the $\mathrm{Aff}(V)$-action, such as the central moments.

We also have the invariant $0$-form $\alpha_0= u$ and $1$-form $du-\lambda_i dx^i$. The latter form vanishes when restricted to $\tilde L$, so we consider instead $\alpha_1 = \lambda_i dx^i$ which also is invariant since $u$, and therefore $du$, is. Since $x^i=-H_{\lambda_i}$ on $\tilde L$, we have $dx^i= x^i_{\lambda_j} d\lambda_j= - H_{\lambda_i \lambda_j} d\lambda_j$. And thus
\[\alpha_1 = x^i_{\lambda_j} \lambda_i d \lambda_j= -H_{\lambda_i \lambda_j} \lambda_i d \lambda_j.\]

\begin{remark}
Notice that there may be situations for which it is desirable to consider proper subgroups only, perhaps because we consider a space $V$ with additional structure. For example, in the case where $V$ is a vector space, it may be more appropriate to consider only the action of $GL(V)$. In this case, the regular $k$th moment $m_k$ will be invariant. 
\end{remark}



\section{Jets and PDEs} \label{Jets}
The appropriate framework for working with differential invariants is the theory of jet spaces. It lets us treat functions, sections of bundles, submanifolds of a fixed dimension and, more generally, solutions of PDEs geometrically. In particular it gives a transparent picture of the algebra of scalar differential invariants. We will use most of this section to fix notation and definitions, sufficient for our use, and refer to the standard literature (for example \cite{KV,KLV,Olver1}) for details. We recommend \cite{Olver2} for a comprehensive introduction the theory of jet spaces and differential invariants, and \cite{Handbook} for a more concise overview. The paper \cite{LieTresse} can be added to either one of these as an updated treatment of the theory of scalar differential invariants. In Section \ref{GroupAction} we show that $\mathrm{Aff}(V)$ is the largest group preserving the metric $\sigma_2$. 

\subsection{Jets}

We have seen that the Legendrian submanifolds in $V \times V^* \times \mathbb R$ can be represented, locally, by a function $I(x)$ on $V$. 


Let $J^k(V)$ denote the space of $k$-jets of functions on $V$. It is a bundle over $V$, and we denote the bundle projection by $\pi_k$. As coordinates on $J^k(V)$ we will use \[x^i,\quad u, \quad u_{x^i},\quad  u_{x^i x^j},\quad  ... , \quad u_{x^{i_1} \cdots x^{i_k}}, \quad i_1 \leq \cdots \leq i_k.\] 
We have 
\[\dim J^k(V)=  n+\binom{n+k}{k}. \]


By identifying  $u_{x^i}$ with $\lambda_i$  we get an identification of $J^1(V)$ with $V \times V^* \times \mathbb R$, and the contact form on $V \times V^* \times \mathbb R$ is identified with the Cartan form on $J^1(V)$. A Legendrian submanifold $\tilde L$ of $J^1(V) \simeq V \times V^* \times \mathbb R$ can be prolonged canonically to an $n$-dimensional submanifold $\tilde L^k \subset J^k(V)$, by requiring that $\tilde L^k$ is an integral manifold of the Cartan distribution on $J^k(V)$.




Alternatively, we may remove information gain from the picture, and consider Lagrangian submanifolds of  $V \times V^*$ with the symplectic form $\omega$. (The information gain may be recovered later, up to an additive constant, by solving the system $I_{x^i}=\lambda_i$.) 

Let $J^k(V \times V^*,n)$ denote the space of $k$-jets of $n$-dimensional submanifolds of $V \times V^*$.  The symplectic form $\omega$ defines a PDE $\mathcal E_1 \subset J^1(V \times V^*,n)$. A submanifold $L \subset V \times V^*$ is a Lagrangian submanifold if and only if its one-jets are contained in $\mathcal E_1$. 

Using coordinates $x^i, \lambda_j$ on $V \times V^*$, a Lagrangian submanifold $L$ is locally determined by $n$ functions $x^i(\lambda)$ for $i=1,...,n$. (In thermodynamics this corresponds to writing the internal energy and volume as functions of temperature and pressure.) 

Since $dx^i=x^i_{\lambda_j} d\lambda_j$ on $L$, the restriction of the symplectic form $\omega =d\lambda_i \wedge d x^i$ to $L$ is given by 
\[\omega|_L= \sum_{i,j=1}^n x^i_{\lambda_j} d \lambda_i \wedge d \lambda_j.\]
Thus, the manifold $L$ is Lagrangian ($\omega|_L=0$) if and only if 
\begin{equation}
F_{ij}=x^i_{\lambda_j}-x^j_{\lambda_i}=0. \label{eq:Fij}
\end{equation}  
The equation (\ref{eq:Fij}) is the coordinate expression for $\mathcal E_1$. We use coordinates 
\[\lambda_i, \quad x^j, \quad x^j_{\lambda_i},  \quad ..., \quad x^j_{\lambda_{i_1} \cdots \lambda_{i_k}}, \quad i_1 \leq \cdots \leq i_k, \]
on $J^k(V \times V^*, n)$. Its dimension is given by
\[\dim J^k(V\times V^*,n) = n+n\binom{n+k}{n}.\]

By differentiating the $\binom{n}{2}$ equations $x^i_{\lambda_j}=x^j_{\lambda_i}$ with respect to the variables $\lambda_1,...,\lambda_n$ we get $n\binom{n}{2}$ additional equations of order two. We add these to the original set of first-order equations, and we denote the corresponding manifold in $J^2(V \times V^*,n)$ by $\mathcal E_2$. Similarly, we get submanifolds $\mathcal E_k \subset J^k(V\times V^*,n)$ for every positive integer $k$, by adding all derivatives (of appropriate order) of the first-order equations. We will also use the notation $\mathcal E_0=J^0(V\times V^*,n) = V \times V^*$.


By counting we easily get the following statement. 
\begin{theorem}
The dimension of $\mathcal E_k$ is given by 
\[\dim \mathcal E_k= \dim J^{k+1}(V)-1 = n+\binom{n+k+1}{n}-1,\]
for $k \geq 0$. 
\end{theorem}
Note that by throwing away the information gain from the picture, we get a natural projection $J^{k+1}(V) \to \mathcal E_k \subset J^k(V \times V^*,n)$, which is reflected in the counting above.

\subsection{The action of the affine group on $J^k(V)$ and $\mathcal E_k$} \label{GroupAction}
In Section \ref{Group} we explained how $\mathrm{Aff}(V)$ acts on $V \times V^* \times \mathbb R$. The action on $V$ induces uniquely an action on $V \times V^* \times \mathbb R$ which preserves the contact form $\theta = du-\lambda_i d x^i$. Another way to say it is that a transformation on $J^0(V)$ induces uniquely a transformation on $J^1(V)$. Moreover, it induces  a transformation on $J^k(V)$ for every positive integer $k$. 

All of these statements can be made in terms of $\mathcal E_k$ in a similar way. Notice that since $\mathrm{Aff}(V)$ preserves the symplectic structure on $V \times V^*$, it preserves the subset $\mathcal E_k \subset J^k(V\times V^*,n)$.

We will illustrate how the prolongation works, by prolonging a vector field on $J^0(V)=V \times \mathbb R$ to a vector field on $J^2(V)$. See for example \cite{KV} for a more general treatment.  Consider the vector field $X=a^i(x) \partial_{x^i}$ on $V \times \mathbb R$. We remind that we want to leave the variable corresponding to the information gain untouched. The unique vector field on $J^1(V)$ which preserves the Cartan distribution and projects to $X$ is given by 
\[ X^{(1)} = a^i(x) \partial_{x^i} - a^s_{x^j}(x) u_{x^s} \partial_{u_{x^j}}.\]
Note that when $a^i$ are affine functions, this corresponds exactly to the $\mathrm{Aff}(V)$-action described in Section \ref{Group}. The prolongation of $X$ to $J^2(V)$ is given by 
\[X^{(2)}= a^i \partial_{x^i} - a^s_{x^j} u_{x^s} \partial_{u_{x^j}} - (D_{x^m}(u_{x^s}) a^s_{x^l} + D_{x^l}(u_{x^s}) a^s_{x^m} + u_s a^s_{x^l x^m})\partial_{u_{x^lx^m}}.\]
In this formula $l\leq m$ is assumed in the summation, and $D_{x^i}$ is the total derivative operator.

Now, let us see which conditions we get on $a^i$ if we require $X$ to preserve the variance $\sigma_2= D_{x^i x^j} (u) dx^i \otimes dx^j$. Here $D_{x^i x^j}= D_{x^j} \circ D_{x^i}$. We compute $L_{X^{(2)}} \sigma_2$. 
\begin{gather*}
L_{X^{(2)}} (\sigma_2) = X^{(2)}(D_{x^i x^j} (u)) dx^i \otimes dx^j\\ + D_{x^i x^j} (u)\left( L_{X^{(2)}}(dx^i) \otimes dx^j+ dx^i \otimes L_{X^{(2)}}(dx^j)\right).
\end{gather*}
We have 
\begin{align*}
X^{(2)}(u_{x^i x^j}) &= -(D_{x^j}(u_{x^s}) a^s_{x^i} + D_{x^i}(u_{x^s}) a^s_{x^j} + u_s a^s_{x^i x^j}), \\
L_{X^{(2)}}(dx^s) &= d(i_{X_f^{(2)}}dx^s)= d(a^s)=a^s_{x^i} dx^i.
\end{align*}
The terms in $L_{X^{(2)}} (\sigma_2)$ that depend on first-order derivatives of $a^i$ cancel, and we see that  $L_{X^{(2)}} (\sigma_2)$ vanishes if and only if all second-order partial derivatives of $a^i$ vanish, implying that $a^i$ are affine functions on $V$. In this sense $\mathrm{Aff}(V)$ is the largest group acting on $V$ which preserves the variance $\sigma_2$.

\begin{theorem} \label{unique}
Let $X= a^i(x) \partial_{x^i}$ be a vector field on $V \times \mathbb R$. Then $X$ preserves the variance $\sigma_2$ (i.e. $L_{X^{(2)}} (\sigma_2)=0$) if and only if $X$ is an affine vector field.
\end{theorem}

\subsection{Differential invariants}
Since $\mathrm{Aff}(V)$ acts on $J^k(V)$, we can look for functions on $J^k(V)$ that are $\mathrm{Aff}(V)$-invariant. 

\begin{definition}
A (scalar) differential invariant of order $k$ is a function on $J^k(V)$ which is constant on $\mathrm{Aff}(V)$-orbits. 
\end{definition}

Let $\tilde L$ be a Legendrian submanifold in $J^1(V) \simeq V \times V^* \times \mathbb R$. A function $\varphi$ on $J^k(V)$ can be restricted to $\tilde L^k \subset J^k(V)$ to give a function $\varphi|_{\tilde L}$ on $\tilde L$. Locally it can be considered as a function on $V$ (or, if one wishes, on $V^*$). If $\varphi$ is a differential invariant of order $k$, it does not mean that the resulting function on $V$ is $\mathrm{Aff}(V)$-invariant, but it means that an $\mathrm{Aff}(V)$-related functions of $\varphi$ on $V$ will depend on $k$-jets of $\tilde L$ in the same way as $\varphi$. 

We mentioned already that the central moments are invariant under the action by $\mathrm{Aff}(V)$. The central moment $\sigma_k$ should be interpreted as a horizontal symmetric $k$-form on $J^k(V)$ which in coordinates takes the form $a_{i_1\cdots i_k} dx^{i_1} \otimes \cdots\otimes  dx^{i_k}$ where $a_{i_1\cdots i_k}$ are functions on $J^k(V)$. This symmetric $k$-form on $J^k(V)$ is $\mathrm{Aff}(V)$-invariant. 

Due to the natural projection $J^{k+1}(V) \to \mathcal E_k \subset J^{k}(V\times V^*,n)$,  any function on $\mathcal E_k$ gives a function on $J^{k+1}(V)$. Essentially, the only thing we loose by considering only functions on $\mathcal E_k$, is the zero-order invariant $u$ corresponding to the information gain. We will mostly take the latter viewpoint in the remaining sections, and therefore in most cases not mention the information gain explicitly in our description of differential invariants. 

We follow \cite{LieTresse} and consider differential invariants that are smooth in the base variables (on $J^0(V)=V \times \mathbb R$), rational in fiber variables of $J^l(V) \to J^0(V)$ for some integer $l$, and polynomial in fiber variables of $J^k(V)  \to J^l(V)$ for $k>l$. In our case the integer $l$ may be taken to be $2$. 
The algebra of differential invariants can be considered as a differential algebra, and in the next section we will show how to construct $n$ independent invariant derivations. By using these, the algebra of differential invariants is generated by a finite number of elements.





\section{Finding differential invariants} \label{Invariants}
In this section we describe the algebra of scalar differential invariants. 
We will construct the scalar invariants by using the central moments $\sigma_k$ which are invariant symmetric $k$-forms on Lagrangian submanifolds of $V \times V^*$, in addition to the invariant $1$-form $\alpha_1$ discussed in Section \ref{Group}.

\begin{remark}
All results in this section may be useful also if we want to consider a subgroup $G\subset \mathrm{Aff}(V)$ (for example corresponding to additional structure on $V$). All scalar and tensorial invariants will be invariant also with respect to $G$. The essential changes in the algebra of scalar differential invariants will occur on the level of third-order invariants, where new invariants will appear. 
\end{remark}

\subsection{From invariant symmetric forms to scalar invariants}
We use $\alpha_1$, $\sigma_2$ and $\sigma_3$ to construct an invariant frame on $L \subset V \times V^*$. When writing $\sigma_k$ in terms of this frame, the coefficients will be scalar differential invariants. More precisely, the invariants are constructed in the following way.
\begin{itemize}
\item The symmetric 2-form $\sigma_2$ is nondegenerate, and may be used to construct a vector $v_1=\sigma_2^{-1}(\alpha_1)$. 
\item By using $\sigma_2$ again we may turn the symmetric 2-form $i_{v_1} \sigma_3$ into a map  $A\colon TL \to TL$. 
\item We use $A$ to define $n-1$ additional vectors: $v_i= A^{i-1} v_1$, for $i=2,...,n$. 
\item The functions $\sigma_k(v_{i_1},...,v_{i_k})$ will be rational, scalar differential invariants.
\end{itemize}
Notice that all these invariants are rational functions, in line with \cite{LieTresse}.

There is only one differential invariant of order $2$ (in addition to the one of order zero). It can be given in coordinates by 
\[\sigma_2^{-1} (\alpha_1, \alpha_1)= \alpha_1(v_1)= x^i_{\lambda_j} \lambda_i \lambda_j. 
\]
This fact is independent of $\dim V$, and can be explained in the following way. 

Consider the action of $GL(V)$ on the 2-jet of the information gain $I(x)$ at $x=0$. We have $j_0^2(I)(x)= I(0)+ a_i x^i+ a_{ij} x^i x^j$. The action by $GL(V)$ preserves the degree of monomials, and we can normalize the quadratic terms to get $j_0^2(I)(x)=I(0)+b_i x^i+ \sum_i (x^i)^2$. Next, we may use the stabilizer $O(V)$ of the quadratic form, and rotate the expression into $j_0^2(I)(x)=I(0)+c x^1+\sum_i (x^i)^2$. There is only one free constant $c$, in addition to $I(0)$, so there is at most one invariant of second order. And we found it.

\begin{theorem}
The algebra of scalar differential invariants is generated by $\sigma_k(v_{i_1},...,v_{i_{k}})$, where $i_1 \leq \cdots \leq i_k$ and $k =2,3,...$. 
\end{theorem}

Notice that 
all of the invariants $\sigma_k(v_{i_1},...,v_{i_{k}})$ are independent for $k \geq 4$. 

When restricted to a Lagrangian submanifold $L \subset V \times V^*$, the set of invariants of the form $\sigma_k(v_{i_1},...,v_{i_{k}})$ will be functions on $L$. However, the resulting set of functions is not completely arbitrary (one can not obtain every set of functions by choosing the appropriate $L$). This is reflected firstly in algebraic relations between these invariants (not all of the invariants are independent). And if we consider the algebra of invariants as a differential algebra, we also have differential syzygies. These two points will be discussed in section \ref{AlgebraicSyzygies}, \ref{DifferentialSyzygies}, respectively. The third, and a more subtle point is that not all sequences of numbers are possible central moments. We will not pay much attention to this point in the current treatment. 

Let us start by counting the number of independent invariants. 

\subsection{The Hilbert function for the differential invariants}
Let $s_k$ count the codimension of an $\mathrm{Aff}(V)$-orbit in $J^k(V)$, in general position, for $k \geq 0$. This number is the same as the transcendence degree of the field of rational scalar differential invariants of order $k$. The Hilbert function for the filtered field of differential invariants is defined as $H_k=s_k-s_{k-1}$ for $k \geq 1$, and $H_0=s_0$. 

Often knowing $H_k$ may be useful when we try to find differential invariants. In this case however, it is clear that we already found all the invariants and we may use our results from the previous section to find the Hilbert function.  We know that $H_0=1, H_1=0, H_2=1$. Moreover since the vector fields $v_i$ depend on 3-jets, it is clear that all the invariants $\sigma_k(v_{i_1},...,v_{i_{k}})$ are independent for $k \geq 4$, thus $H_k=\binom{n+k-1}{k}$ for $k \geq 4$. 

The only formula which can not be read off directly from the previous section is that of $H_3$. But since $\dim \mathrm{Aff}(V)=n^2+n$, and $\dim J^3(V)= n+\binom{n+3}{3}$, we get 
\[s_3=\dim J^3(V)-\dim \mathrm{Aff}(V)= \frac{n^3+11n+6}{6},\]
and $H_3=s_3-s_2=s_3-2$. 

\begin{theorem}
The Hilbert function for the field of differential invariants in $J^k(V)$ is given by
\begin{gather*}
H_0=1, \qquad H_1= 0, \qquad H_2=1, \qquad  H_3=\frac{n^3+11n-6}{6}  \\
H_k = \binom{n+k-1}{k}, \qquad k \geq 4. 
\end{gather*}
\end{theorem}
Note that for $n=2$, the formula for $H_3$ coincides with that for $H_k$, as both gives $4$. 


We define the Poincaré function corresponding to the Hilbert function by the series $P(z)=\sum_{k=0}^\infty H_k z^k$ which converges to a rational function for $|z|<1$. 
\begin{theorem}
The Poincaré function is given by 
\[P(z)= (1-z)^{-n}- \frac{z}{2} \left((n-1)(n-2)z^2+(n+2)(n-1)z-2n    \right).\]
\end{theorem}

\subsection{Algebraic relations} \label{AlgebraicSyzygies}
It is clear that the invariants $\sigma_k(v_{i_1},...,i_{k})$ are independent for $k \geq 4$. But in order to get the complete picture, we will also find the relations between the coefficients of $\sigma_2$ and $\sigma_3$. First of all, due to the construction of the frame $\{v_1,...,v_n\}$, we have 
\[i_{v_{i}}\sigma_2 = i_{v_{i-1}} i_{v_1} \sigma_3.\] 
Note how this is consistent with the fact that we have only one second-order invariant $\sigma_2(v_1,v_1)$. It also follows that there will be relations between $\sigma_3(v_i,v_j,v_k)$: 
\[\sigma_3(v_1,v_i,v_j)=\sigma_3(v_1,v_{i+1},v_{j-1}).\] In two dimensions this holds trivially, since $\sigma_3$ is symmetric. In three dimensions we get the additional relation
\[\sigma_3(v_1,v_1,v_3)=\sigma_3(v_1,v_2,v_2).\] 
In four dimensions we also have 
\[\sigma_3(v_1,v_1,v_4)=\sigma_3(v_1,v_2,v_3), \quad  \sigma_3(v_1,v_2,v_4)=\sigma_3(v_1,v_3,v_3).\]
For  $\dim V=n \geq 3$ we get $\binom{n-1}{2}$ relations between the $\binom{n+3-1}{3}$ components of $\sigma_3$. The difference is $\frac{n^3+11n-6}{6}$ which is exactly $H_3$ from the previous section.


\subsection{The differential algebra of differential invariants}
Above, the algebra of differential invariants was generated by an infinite number of elements. However, if we consider it instead as a differential algebra it will be finitely generated (\cite{LieTresse}). The vectors $v_1,...,v_n$ can be considered as invariant derivations. They act on the differential invariants and, together with a finite set of differential invariants, they generate the whole field. 

\begin{theorem}
The algebra of scalar differential invariants is generated by the invariant derivations $v_1,...,v_k$ and the scalar invariants $\sigma_2(v_1,v_1)$, $\sigma_3(v_i,v_j,v_k)$, and $\sigma_4(v_i,v_j,v_k,v_l)$. 
\end{theorem}

This is not a freely generated algebra; there are differential syzygies among the generators. We will compute the syzygies in the simplest case, when $\dim V=2$. 

\subsection{Differential syzygies for $\dim V=2$} \label{DifferentialSyzygies}

Let us use the notation
{\small \begin{gather*}
I_{21}=\sigma_2(v_1,v_1),\;I_{22}=\sigma_2(v_1,v_2),\;I_{23}=\sigma_2(v_2,v_2),\; I_{31}=\sigma_3(v_1,v_1,v_1),\\ I_{32}= \sigma_3(v_1,v_1,v_2), \quad  I_{33}=\sigma_3(v_1,v_2,v_2),\quad  I_{34} = \sigma_3(v_2,v_2,v_2).
\end{gather*}}

We have $I_{22}=I_{31}$ and $I_{23}=I_{32}$. In order to write the differential syzygies in relatively compact form, it will be useful to have the following definitions:
\begin{gather*}
J_1=\frac{I_{21} I_{33}-I_{22} I_{32}}{I_{21}I_{23}-I_{22}^2}, \qquad J_2=\frac{I_{22} I_{33}-I_{23} I_{32}}{I_{21}I_{23}-I_{22}^2}, \\
J_3=\frac{I_{21} I_{34}-I_{22} I_{33}}{I_{21}I_{23}-I_{22}^2}, \qquad J_4=\frac{I_{22} I_{34}-I_{23} I_{33}}{I_{21}I_{23}-I_{22}^2}
\end{gather*}
When $\dim V=2$, the third-order invariants are sufficient to generate the whole algebra. Thus we consider the generators 
\[I_{21}, \quad I_{31}, \quad I_{32}, \quad I_{33}, \quad I_{34}.\]

Let us first consider the invariant derivatives of the four third-order invariants. Since $H_4=5$, the invariant derivations give at most 5 new invariants that are independent of the previous ones. It is easily verified that this upper bound is obtained, meaning that there must be three differential syzygies. They are given by 
\begin{gather*}
2 v_2(I_{31})-v_1(I_{32})-I_{33}-2I_{32}=0, \\
\Big(v_2(I_{32})-2 v_1(I_{33})+2 J_1 v_1(I_{32})-4 J_2 v_1(I_{31})+I_{34} \\-2 J_1 I_{33}+4 J_2 I_{32}+(6 J_2-2 J_1^2) I_{31}+2J_1 J_2 I_{21}\Big) =0,\\
\Big( v_2(I_{33})-v_1(I_{34})-J_1 v_2(I_{32})+(2 J_2+3 J_3) v_2(I_{31}) \\ -3 J_4 v_1(I_{31})-(2 J_1+4) I_{34}+(I_{33}-2 I_{32}) J_2\Big) = 0.
\end{gather*}
If we differentiate $I_{21}$ we get 
\[v_1(I_{21})=2 I_{21}+I_{31}, \qquad v_2(I_{21})=2I_{31}+I_{32}.\]
The derivations $v_1, v_2$ satisfy the commutation relation
\begin{align*}
[v_1,v_2]&=\left(\frac{(I_{33}-I_{42})I_{22}-I_{32} (I_{23}-I_{41})}{I_{21} I_{23} -I_{22}^2}-3I_{21}\right) v_1\\ &+\left(\frac{(I_{42}-I_{33})I_{21}-I_{31}(I_{41}-I_{23})}{I_{21} I_{23} -I_{22}^2}+1\right) v_2. 
\end{align*}

%
%

\section{Thermodynamics of gases} \label{Gases}
We will now see what differential invariants appear in the context of gases in thermodynamic equilibrium. Consider the thermodynamic space with variables $p,T,e,v,s$ corresponding to pressure, temperature, internal energy, volume and entropy. The entropy is related to the information gain $I$ by the formula $dI=-ds$. By aligning the one-form $\theta=dI-\lambda_i dx^i$ with the fundamental thermodynamic relation $-ds+T^{-1} de+pT^{-1} dv=0$, we see that we can get the standard thermodynamics of gases from the measurement of a point $(e,v) \in V$, and the principle of minimal information gain. The relationship between $p,T,e,v$ and $x^1,x^2,\lambda_1,\lambda_2$ is 
\[x_1=e, \qquad x_2=v, \qquad \lambda_1=-T^{-1},\qquad \lambda_2=-pT^{-1}  .\]

We will suppress the information gain, or entropy, from the picture, and consider a thermodynamic state as a Lagrangian submanifold of $V\times V^*$ on which the symplectic form 
\[\omega =d\theta=\frac{1}{T^2} (de \wedge dT+p dv \wedge dT+T  dp\wedge dv)\]
vanishes. We will assume that $T\neq 0$ throughout.  

Similarly as above, we let the Lagrangian submanifold $L \subset V \times V^*$  be given by two functions 
 $e(T,p),v(T,p)$. Restricting $\omega$ to such a submanifold gives 
\[\omega|_L= \frac{1}{T^2} (e_p+p v_p+T v_T) dp \wedge dT,\] 
implying that $L$ is Lagrangian if and only if \[F=e_p+T v_T+pv_p=0.\] The differential equation $F=0$ determines the submanifold $\mathcal E_1$ in $J^1(V^* \times V,n)$. 


\subsection{Group action}
In the (nonlinear) coordinates $T,p,e,v$, the action of $\mathrm{Aff}(V)$ on $V\times V^*$ looks slightly different. The corresponding Lie algebra of vector fields is spanned by the six vector fields
\begin{align*}
\partial_e,\,\partial_v,\, \partial_p-v \partial_e,\, e\partial_e+T \partial_T+p\partial_p,\, v \partial_v-p\partial_p,\, e\partial_v+ Tp\partial_T+p^2\partial_p.
\end{align*}
Orbits in general position in $\mathcal E_1$ are six-dimensional. The subset on which the orbit dimension decreases is given by $T(e_p v_T-e_T v_p)=0$. Positive definiteness of $\sigma_2$ implies positivity of the left-hand side of this equation (see the end of Section \ref{GasInvariants}).


One can ask if it is natural to consider arbitrary affine transformation on points $(e,v)$. Here we want to make as few assumptions as possible when it comes to choosing Lie subgroup of $\mathrm{Aff}(V)$, and we will consider only two choices. The first obvious choice is the biggest group possible, which is the one we have been studying so far. We will see that the simplest invariant $\sigma_2(v_1,v_1)$ is what is known in thermodynamics as the heat capacity. 

The second choice of group will be based on recent results concerning the thermodynamics of fluids. In \cite{DLTviscid2D} they find the symmetries of compressible viscid fluids, namely symmetries of the Navier-Stokes equations (including an equation for heat transfer). Some of the symmetries are purely geometrical: translations, rotations and Galilean transformations. In addition there are some ``thermodynamic'' symmetries that act on thermodynamic variables. The intersection between these thermodynamic symmetries and our group of affine transformation is two-dimensional. It contains exactly the scalings on $e$ and $v$, respectively. Its infinitesimal generators are 
\[e\partial_e+T \partial_T+p\partial_p,\qquad v \partial_v-p\partial_p .\]

\subsection{Differential invariants with respect to $\mathrm{Aff}(V)$} \label{GasInvariants}
The algebra of differential invariants for gases was completely described in section \ref{Invariants}. We found both generators and syzygies. 
Those results can be directly applied to our current case of gases. Here we have worked, with the help of the DifferentialGeometry and JetCalculus packages in Maple, to find invariants whose expressions are simpler in the current choice of coordinates.

First of all, we have the following invariant derivations:
\[ \nabla_1=-TD_T, \qquad \nabla_2 = \frac{T v_T D_T+(e_T+pv_T)D_p}{(e_T v_{TT}-e_{TT} v_T)T}\]
And we remember from the previous section that there should be (in addition to the information gain) one second-order invariant and $k+1$ new invariants of order $k$ for $k \geq 3$. 
\begin{theorem}
The field of third-order differential invariants is generated by the following differential invariants:
\begin{gather*}
 I_2=pv_T+e_T,\quad I_{31}=-(pv_{TT} + e_{TT}),\quad I_{32}=\frac{(e_T v_{TT}-e_{TT} v_T)^2 T^3}{e_p v_T-e_T v_p}, \\ I_{33}=\frac{\left(2 T v_T I_2 v_{TT}+I_2^2 v_{Tp}+v_T^2 (I_2+I_{31})\right)T}{(e_p v_T-e_T v_p)},\\
   I_{34}=\tfrac{T^2\left(3 T^2 v_T^2 I_2 v_{TT}+3T v_T I_2^2 v_{Tp}+I_2^3 v_{pp} +T v_T^3 I_{31}+4T v_T^3 I_2+3v_T v_p I_2^2\right)(e_T v_{TT}-e_{TT} v_T)}{(e_p v_T-e_T v_p)^2} 
\end{gather*}
\end{theorem}
These are not exactly the invariants we used above (although they obviously generate the same field), but $I_2= \sigma_2(v_1,v_1)$. 
We notice that this second-order differential invariant is exactly what is known in thermodynamics as the heat capacity (at constant pressure). Thus the concept of heat capacity is given to us automatically if we consider the action of the affine group on $V\times V^*$ (a subgroup of the affine group will lead to even more invariant quantities). 

Let us follow the tradition in geometry of considering manifolds with constant curvature, and find the thermodynamic states with constant heat capacity. 
\begin{example}
The thermodynamic states of constant heat capacity are solutions of the system \[F=e_p+T v_T+p v_p=0,\qquad I_2=pv_T+e_T=C \] for $C \in \mathbb R$. They are given by 
\[e=f_1(p) T -f_2'(p)p^2, \qquad v=\frac{(C-f_1(p))T}{p}+ f_2'(p) p+f_2(p). \] 
\end{example}

By using the invariant derivations $\nabla_1$ and $\nabla_2$ we can generate the algebra of differential invariants.
\begin{theorem}
The algebra of scalar differential invariants is generated by the invariant derivations $\nabla_1$ and $\nabla_2$, together with the differential invariants $I_2,I_{32},I_{33},I_{34}$. 
\end{theorem}
Notice that $I_{31}=\nabla_1(I_2)$. 

The second central moment is given by 
\[\sigma_2=\frac{1}{T} (I_2 dT^2-2v_T dT dp- v_p dp^2).\] In the frame $\nabla_1,\nabla_2$ it takes diagonal form: 
\[ \sigma_2(\nabla_1,\nabla_1)=I_2, \qquad \sigma_2(\nabla_1,\nabla_2)=0,\qquad \sigma_2(\nabla_2, \nabla_2)= I_2/I_{32}\]
Positive definiteness of $\sigma_2$ implies $I_1>0$ and $(e_p v_T-e_T v_p) T >0$.


\subsection{Invariants with respect to a two-dimensional subgroup}

Now, let us consider the two-dimensional group which scales $e$ and $v$, respectively: $(e,v) \mapsto (t e, s v)$, $t,s \in \mathbb R \setminus \{0\}$. The corresponding Lie algebra is spanned by 
\[X=e\partial_e+T \partial_T+p\partial_p,\qquad Y=v \partial_v-p\partial_p .\]
In this case we get 5 invariants of order two. 

\begin{theorem}
The field of second-order differential invariants is generated by
\begin{gather*}
e/T, \qquad pv/T, \qquad e_T\\
e_T+p v_T, \qquad e_T-e_p \frac{v_T}{v_p}.
\end{gather*}
\end{theorem}
The last two invariants are the heat capacity at constant pressure and at constant volume, respectively. The derivations $T D_T$ and $p D_p$ are invariant, and by using them we can generate the algebra of differential invariants. 

\begin{theorem}
The algebra of differential invariants is generated by the following two differential invariants of order 1 and two invariant derivations:
\[J_1=e/T, \qquad J_2=pv/T, \qquad \nabla_1 = T D_T, \qquad \nabla_2=p D_p\]
The differential syzygy is
\[\nabla_2(J_1)+\nabla_1(J_2)+\nabla_2(J_2)=0 .\]
\end{theorem}

We can also generate the algebra by using Tresse derivatives $\hat \partial_1, \hat \partial_2$ with respect to the pair $J_1,J_2$.  They are defined by $\hat \partial_i(J_j)=\delta_{ij}$, and therefore play the role of partial derivatives with respect to $J_1$ and $J_2$. 
We denote the remaining second-order invariants by \[K_1=e_T, \qquad K_2=p v_T, \qquad K_3=p^2 v_p/T.\] 
Clearly, $J_i,K_j,\hat \partial_i$ generate the algebra of differential invariants. There are two first-order syzygies:
 \begin{align*}
0= &\Big((K_2+K_3) \hat \partial_1(K_1)-(J_2+K_3) \hat \partial_2(K_1)\\ &+(J_1-K_1+K_2+K_3) \hat \partial_1(K_2)-(K_2+K_3) \hat \partial_2(K_2)\Big) \\ 
0= &\Big(-(K_2+K_3) \hat \partial_1(K_2)+(J_2+K_3) \hat \partial_2(K_2)\\ &+(J_1-K_1) \hat \partial_1(K_3)+(J_2-K_2) \hat \partial_2(K_3)-K_2-K_3\Big)
\end{align*}
When written in terms of the Tresse derivatives, the syzygies are easily interpreted as differential equations. In this case there are two equations on three functions of two variables. They are often called the quotient or factor equations. Their solutions give allowed relations between the invariants, when restricted to a Lagrangian submanifold. Equivalent Lagrangian manifolds (in the sense of the group action) will give rise to the same relations between the invariants. 

Note that the Lagrangian manifolds on which the first two invariants $J_1,J_2$ are constant are exactly the ideal gases:
\[e=C_1 T, \qquad pv=C_2 T\] 
In this way ideal gases are singled out as special states of ``constant curvature'', with respect to the action of the two-dimensional Lie group. 

This follows as a consequence of the fact that these states are exactly the invariant Lagrangian submanifolds with respect to the two-dimensional Lie group. It can be showed in the following way.
A Lagrangian submanifold is given by $d\theta|_L=0$. The additional requirement that $L$ is invariant is given by 
\[ (i_X d\theta)|_L=0, \qquad (i_Y d\theta)|_L=0.\] 
In other words, the vector fields $X$ and $Y$ are tangent to $L$. 
Assuming that $L$ is given by $e(T,p),v(T,p)$, these three can be written as  five differential equations on $e$ and $v$:
\begin{align*}
T v_T+p v_p+e_p = 0, \quad e=T e_T, \quad e_p=0, \quad v=T v_T, \quad v=-pv_p
\end{align*}
The solutions are exactly the ideal gases.

\vspace{0.5cm}

\noindent \textbf{Acknowledgements:} This project was supported by the Czech Science Foundation (GA\v{C}R no. 19-14466Y).


\end{document}